\documentclass[aps, twocolumn,superscriptaddress, 
amsmath,amssymb,reprint,numbers,noeprint
]{revtex4-1}

\usepackage[toc,page]{}
\usepackage{siunitx}
\usepackage{svg}
\usepackage{gensymb}
\usepackage{graphicx}
\usepackage{dcolumn}
\usepackage{bm}
\begin{document}

\preprint{APS/123-QED}

\title{\textcolor{black}{Wide-field dynamic magnetic microscopy using double-double quantum driving of a diamond defect ensemble}}




\author{Zeeshawn Kazi}
\email{zeeshawn@uw.edu}
\affiliation{University of Washington, Physics Department, Seattle, WA, 98105, USA}%

\author{Isaac M. Shelby}%
\affiliation{University of Washington, Physics Department, Seattle, WA, 98105, USA}%

\author{Hideyuki Watanabe}
\affiliation{Nanoelectronics Research Institute,
National Institute of Advanced Industrial Science and Technology,
Tsukuba Central 2, 1-1-1 Umezono, Tsukuba, Ibaraki 305-8568, Japan}
\author{Kohei M. Itoh}
\affiliation{Spintronics Research Center, Keio University, 3-14-1 Hiyoshi, Kohoku-ku, Yokohama 223-8522, Japan}
\affiliation{School of Fundamental Science and Technology, Keio University, 3-14-1 Hiyoshi, Kohoku-ku, Yokohama 223-8522, Japan}
\author{Vaithiyalingam Shutthanandan}
\affiliation{Environmental and Molecular Sciences Laboratory, Pacific Northwest National Laboratory, Richland, WA, USA}
\author{Paul A. Wiggins}
\affiliation{University of Washington, Physics Department, Seattle, WA, 98105, USA}%
\affiliation{University of Washington, Bioengineering Department, Seattle, WA, 98105, USA}
\author{Kai-Mei C. Fu}

\affiliation{University of Washington, Physics Department, Seattle, WA, 98105, USA}%
\affiliation{University of Washington, Electrical and Computer Engineering Department, Seattle, WA, 98105, USA}

\begin{abstract}




Wide-field magnetometry can be realized by imaging the optically-detected magnetic resonance of diamond nitrogen vacancy (NV) center ensembles. However, \textcolor{black}{NV ensemble} inhomogeneities significantly limit the magnetic-field sensitivity of these measurements. \textcolor{black}{We demonstrate a double-double quantum (DDQ) driving technique to facilitate wide-field magnetic imaging of dynamic magnetic fields at a micron scale.} DDQ imaging employs four-tone radio frequency pulses to suppress inhomogeneity-induced variations of the NV resonant response. \textcolor{black}{As a proof-of-principle, we use the DDQ technique to image the dc magnetic field produced by individual magnetic-nanoparticles tethered by single DNA molecules to a diamond sensor surface. This demonstrates the efficacy of the diamond NV ensemble system in high-frame-rate magnetic microscopy, as well as single-molecule biophysics applications.}

\end{abstract}

\maketitle

\section*{Introduction}
The success of the negatively-charged nitrogen-vacancy (NV) center in diamond \textcolor{black}{for magnetic field sensing} is due to a \textcolor{black}{powerful combination of characteristics:} a long spin-coherence time, the ability to perform optically-detected magnetic resonance (ODMR) \textcolor{black}{spectroscopy} at room temperature, and a solid-state host environment which facilitates sample-sensor integration~\cite{Abe2018Tutorial:Magnetometry, Hong2013NanoscaleDiamond,Schirhagl2014Nitrogen-VacancyBiology,Balasubramanian2009UltralongDiamond}. One exciting emerging application is magnetic field imaging using \textcolor{black}{a diamond sensor comprising a diamond substrate with a thin layer of NV centers fabricated at the top surface}~\cite{Levine2019PrinciplesMicroscope,Tetienne2017QuantumGraphene,Barry2016OpticalDiamond,Kehayias2019ImagingCenters,Schlussel2018Wide-FieldDiamond}. In this NV ensemble imaging system, it is critical that the inhomogeneities across the \textcolor{black}{imaging area} are eliminated in order to reach the \textcolor{black}{sensitivity} floor given by \textcolor{black}{a} single-pixel resonance curve. To this end, the NV community has demonstrated many sensor growth and fabrication methods that increase NV density and homogeneity~\cite{Acosta2009DiamondsApplications,Ohno2014Three-dimensionalImplantation,Osterkamp2019EngineeringDiamond,Eichhorn2019OptimizingSensing,Tetienne2018SpinDiamond,Kleinsasser2016}, as well as quantum control methods that suppress external\textcolor{black}{-}field dependence and increase ensemble \textcolor{black}{spin} coherence~\cite{Bauch2018UltralongControl,Myers2017Double-QuantumCenters,Mamin2014MultipulseCenters,DeLange2012ControllingDiamond}. In this paper, we expand on these techniques to present a quantum control method compatible with \textcolor{black}{high-frame-rate} magnetic-field imaging. Our method, double-double quantum (DDQ) driving, mitigates \textcolor{black}{inhomogeneities} caused by variations of the \textcolor{black}{NV} resonance curve.

\begin{figure*}
  \centering
   \includegraphics[width=2.0\columnwidth]{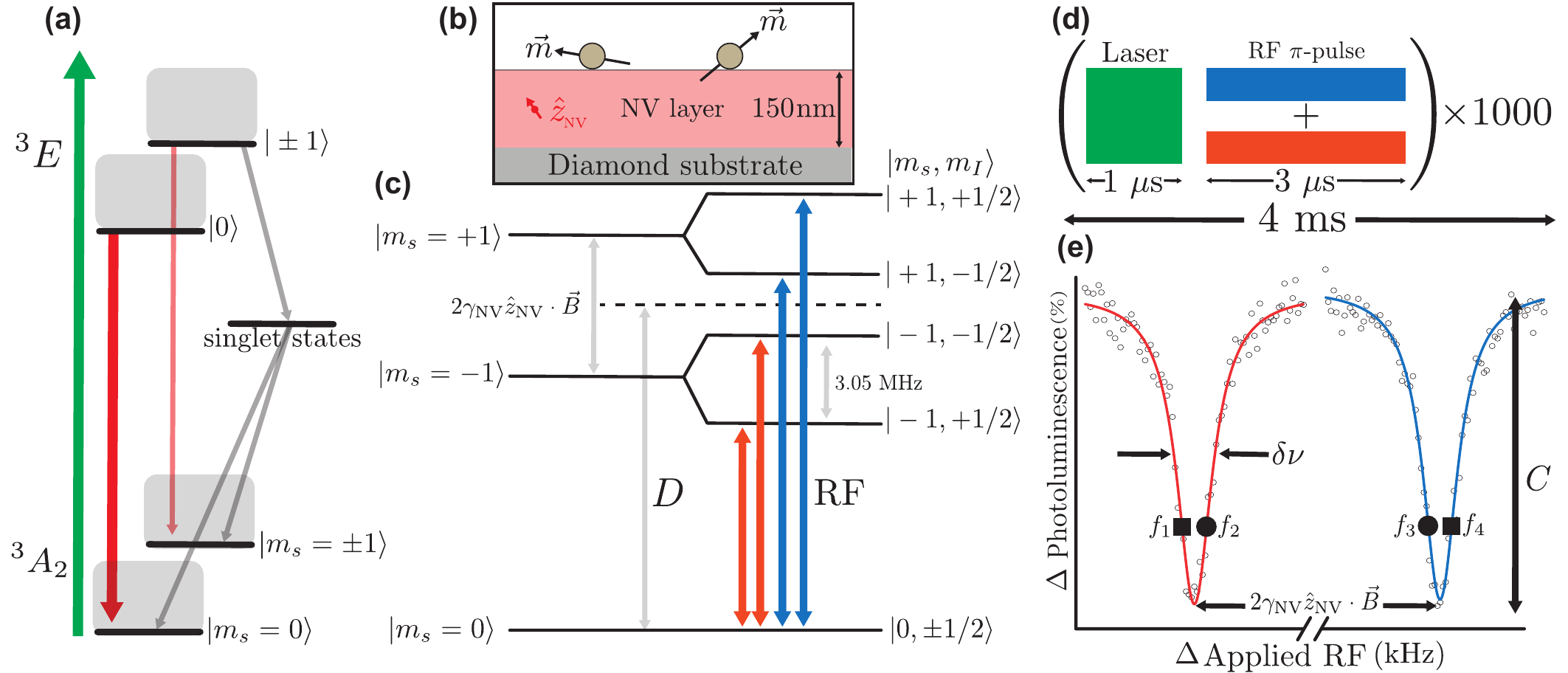}

\caption{Wide-field pulsed magnetic imaging using an NV ensemble. (a) NV electronic energy level diagram showing the ground, excited spin states ($|m_s =0\rangle$,$|m_s = \pm1\rangle$), singlet states, optical excitation (green arrow), emitted photoluminescence (red arrows), and spin-selective, non-radiative inter-system-crossing (gray arrows). (b) Schematic showing 50~nm ferromagnetic nanoparticles adhered to the diamond surface. The magnetic moments of the particles are oriented randomly. The 150~nm NV layer (pink) is fabricated on top of the diamond substrate (\textcolor{black}{gray}), and a single NV pointed along the (111) orientation is shown (red). (c) NV ground state energy level diagram showing the zero-field splitting ($D$), Zeeman splitting of the $|m_s = \pm1\rangle$ states ($2\gamma_{\text{NV}}\hat{z}_{\text{NV}}\cdot\vec{B}$), and $^{15}$N-NV hyperfine splitting (3.05 MHz). RF excitation (orange and blue arrows) rotates the NV spin between the $|m_s = 0\rangle$ and $|m_s = \pm1\rangle$ states. Two-tone RF excitation is simultaneously driven over the two $^{15}$N-NV hyperfine transitions to produce a single combined resonance for each NV electron spin state ($|m_s = \pm1\rangle$). (d) Laser and multi-tone RF $\pi$-pulses are repeated throughout the camera exposure to facilitate wide-field imaging. (e) A Lorentzian-shaped reduction in NV PL is observed when a RF scan is performed through \textcolor{black}{each} spin-transition frequency. The resonances are Zeeman split by the external magnetic field. The outer (inner) inflection points $f_1,f_4$ ($f_2,f_3$) are denoted by black squares (circles). For the resonance curves shown, the FWHM linewidth $\delta\nu = 300\ \text{kHz}$ and fractional optical contrast $C=0.03$, with optical pulse = \SI{500}{\nano\second}, RF pulse = \SI{3500}{\nano\second}, photon collection rate = $1.1\times10^{7} $\SI{}{\hertz} from a \SI{1}{\micro\meter^{2}} pixel (\SI{0.15}{\micro\meter^{3}} voxel), and integration time per data point = \SI{144}{\milli\second}.} 
\label{fig:NVexplanation} 
\end{figure*}

In NV ensemble ODMR, magnetic fields can be imaged by characterizing the photoluminescence of optically-excited NV centers \textcolor{black}{after} \textcolor{black}{probing} the NV spin \textcolor{black}{ground} states with radio-frequency (RF) \textcolor{black}{$\pi$}-pulses [Fig.~\ref{fig:NVexplanation}]. The resonant RF depends on magnetic field because of the Zeeman splitting of the \textcolor{black}{$|m_s = \pm1\rangle$} NV electronic spin states~\cite{Abe2018Tutorial:Magnetometry}. However, in addition to sensitivity to magnetic field, the resonances are also perturbed by inhomogeneities in electric field, temperature, and crystal strain~\cite{Dolde2011Electric-fieldSpins,Mittiga2018ImagingDiamond,Acosta2010TemperatureDiamond,Broadway2019MicroscopicSensors}. Due to the symmetry of the NV center, these \textcolor{black}{non-magnetic} perturbations affect the two NV electron spin resonances \textcolor{black}{($|m_s = 0\rangle\leftrightarrow|m_s = \pm1\rangle$)} in the same way and can be eliminated by characterizing both resonances~\cite{Mamin2014MultipulseCenters,Fang2013High-SensitivityCenters}. 

For wide-field magnetic imaging of time-varying fields, \textcolor{black}{the} full characterization of the resonance curves \textcolor{black}{is too slow to capture the magnetic-field dynamics in many applications}. In \textcolor{black}{these dynamic applications}, shifts of one resonance curve can instead be mapped to changes in emitted NV photoluminescence (PL) intensity by applying single-frequency RF excitation~\cite{Pham2011MagneticEnsembles,McCoey2019RapidMicrobeads,Wojciechowski2018Camera-limitsSensor}. This \textcolor{black}{``single-quantum" (SQ)} imaging modality enables partial reconstruction of the local magnetic field with a higher frame-rate. The double-quantum (DQ) \textcolor{black}{modality}, which drives both \textcolor{black}{NV electron} spin transitions simultaneously by applying a two-tone RF pulse, eliminates pixel-to-pixel \textcolor{black}{non-magnetic} perturbations of the transition resonant frequencies~\cite{Mamin2014MultipulseCenters,Fang2013High-SensitivityCenters}. However, variations of the shape of the resonance curve also \textcolor{black}{cause changes in the emitted PL intensity and therefore generate spurious contrast which limits the magnetic sensitivity}. These variations arise from inhomogeneities in NV and other paramagnetic spin densities as well as external fields~\cite{Bauch2018UltralongControl}, \textcolor{black}{and as shown in this work, can severely limit the utility of the DQ modality for wide-field imaging applications}. By expanding to a four-tone DDQ driving scheme, we suppress \textcolor{black}{anomalous contrast due to resonance-}curve-shape variations \textcolor{black}{pixel-by-pixel} across \textcolor{black}{the} field of view. \textcolor{black}{This enables high-frame-rate imaging of time-dependent magnetic fields. We first demonstrate the SQ, DQ, and DDQ imaging modalities by imaging static fields, and show the DDQ signal is linearly proportional to the magnetic field projection along the NV symmetry axis. We then use the DDQ driving technique to image the dynamic magnetic field produced by a ferromagnetic nanoparticle tethered by a single DNA molecule to the diamond sensor surface.}

\section*{Experimental Methods}
The wide-field NV magnetic particle imaging (magPI) platform used in this work utilizes a diamond sensor with a near-surface, high density NV ensemble [Fig.~\ref{fig:NVexplanation}(b)]. A \SI{150}{\nano\meter} $^{15}$N doped, isotope-purified (99.999$\%$ $^{12}$C) layer was grown by chemical vapor deposition on an electronic-grade diamond substrate (Element Six). The sample was implanted with 25~keV He$^{+}$ at a dose of $5 \times 10^{11}$~ions/cm$^{2}$ to form vacancies, followed by a vacuum anneal at 900~$\degree$C for \SI{2}{\hour} for NV formation and an anneal in $\text{O}_{2}$ at 425~$\degree$C for \SI{2}{\hour} for charge state stabilization~\cite{Kleinsasser2016}. The resulting ensemble has NV density of $1.7\times10^{16}$~cm$^{-3}$ and ensemble spin coherence time $T^{*}_{2} = 2.5$~\SI{}{\micro\second} \textcolor{black}{(further details in Supplemental Material~\cite{DDQSM2020})}.

The NV electronic structure and optical and RF control are summarized in Fig.~\ref{fig:NVexplanation}. A 532 nm laser pulse is used to optically pump the NV ensemble into the $|m_s = 0\rangle$ triplet ground state. RF excitation drives transitions from this ground state into the $|m_s = \pm1\rangle$ spin states. Optical excitation from the $|m_s = \pm1\rangle$ states results in a reduction of PL intensity due to a spin selective, nonradiative inter-system-crossing~\cite{Gali2019AbDiamond}. This relaxation provides the spin-dependent PL contrast and initialization into the $|m_s=0\rangle$ state. Monitoring the emitted PL as a function of RF enables measurement of ODMR for each NV ground state spin transition [Fig.~\ref{fig:NVexplanation}(e)]. 

From a Lorentzian-shaped resonance curve [see Fig.~\ref{fig:NVexplanation}(e)], the \textcolor{black}{volume-normalized,} shot-noise-limited dc magnetic sensitivity is given by
\begin{equation}
  \textcolor{black}{\eta_{SQ}^{V} \approx \frac{1}{\gamma_{\text{NV}}}\frac{\delta\nu}{C}\sqrt{\frac{V}{N}}},
  \label{equation:sensitivity}
\end{equation}
where $C$ is the optical contrast (fractional depth of the resonance curve), $\delta\nu$ is the full width at half maximum (FWHM) linewidth, $\gamma_{\text{NV}}$ is the NV gyromagnetic ratio ($28$ \SI{}{\mega\hertz}/\SI{}{\milli\tesla}), $N$ is the photon detection rate \textcolor{black}{and $V$ is the collection volume}~\cite{Abe2018Tutorial:Magnetometry}. In the magPI platform, the average \textcolor{black}{volume normalized} sensitivity \textcolor{black}{$\eta_{SQ}^{V}~\approx~ $\SI{31}{\nano\tesla\ \hertz^{-1/2}\ \micro\meter^{3/2}}}. All three sensor parameters $C, \delta\nu$, and $N$ vary across the imaging field of view. 

Optical power broadening of the resonance curve is eliminated by using pulsed excitation in which optical and RF fields are applied separately~\cite{Dreau2011AvoidingSensitivity}. Optical pulses and RF $\pi$-pulses (both \SI{}{\micro\second}-scale) are applied to the ensemble repeatedly to fill a sCMOS camera exposure (\SI{}{\milli \second}-scale). Each camera exposure is taken with a single set of RFs with a fixed pulse duration [Fig.~\ref{fig:NVexplanation}(d)], enabling pulsed NV ensemble control and readout with wide-field camera exposure times~\cite{Steinert2013MagneticResolution}.

RF excitation is delivered via a broadband microwave antenna with transmission resonance at the NV zero-field-splitting $D$~\cite{Sasaki2016BroadbandDiamond}. Each RF applied is mixed to create two equal tones separated by \SI{3.05}{\mega\hertz}, which enables simultaneous driving of the two $^{15}$N-NV hyperfine transitions~\cite{Doherty2012TheoryDiamond} [Fig.~\ref{fig:NVexplanation}(c)] and produces one combined resonance for each $|m_s=0\rangle\leftrightarrow|m_s = \pm1\rangle$ transition [Fig.~\ref{fig:NVexplanation}(e)]. Samarium cobalt ring magnets (SuperMagnetMan) are used to apply a \SI{1}{\milli\tesla} static external magnetic field along the (111)~NV orientation ($\hat{z}_{\text{NV}}$). More details about the experimental setup can be found in \textcolor{black}{the Supplemental Material.}

\textcolor{black}{We first use the magPI platform} to image the static dipolar magnetic field produced by a \SI{50}{\nano\meter} dextran coated $\text{Co}\text{Fe}_{\text{2}}\text{O}_{\text{4}}$ ferromagnetic nanoparticle (micromod Partikeltechnologie) deposited and dried onto the diamond sensor surface [Fig. 1(b)]. These bio-compatible particles produce nano-scale magnetic fields which lie in the dynamic range of the NV sensing ensemble defined by $\delta\nu$ of the resonance curves. For other imaging applications, the sensor dynamic range can be increased at the expense of magnetic sensitivity by RF broadening the resonance curve.

\textcolor{black}{We then demonstrate dynamic magnetic imaging using the DDQ technique on a tethered-particle-motion (TPM) assay~\cite{Finzi1995}. \SI{500}{\nano\meter} streptavidin coated ferromagnetic nanoparticles (micromod Partikeltechnologie 05-19-502) were tethered by 940~bp single DNA molecules to the diamond surface. The diamond-DNA-particle tethering protocol follows Ref. \cite{Kovari2018}. Fluid flowed through the sample chamber alters the orientation of the nanoparticle magnetic moment, and the changing magnetic field is imaged at a high frame-rate.} 


\begin{figure*}
   \includegraphics[]{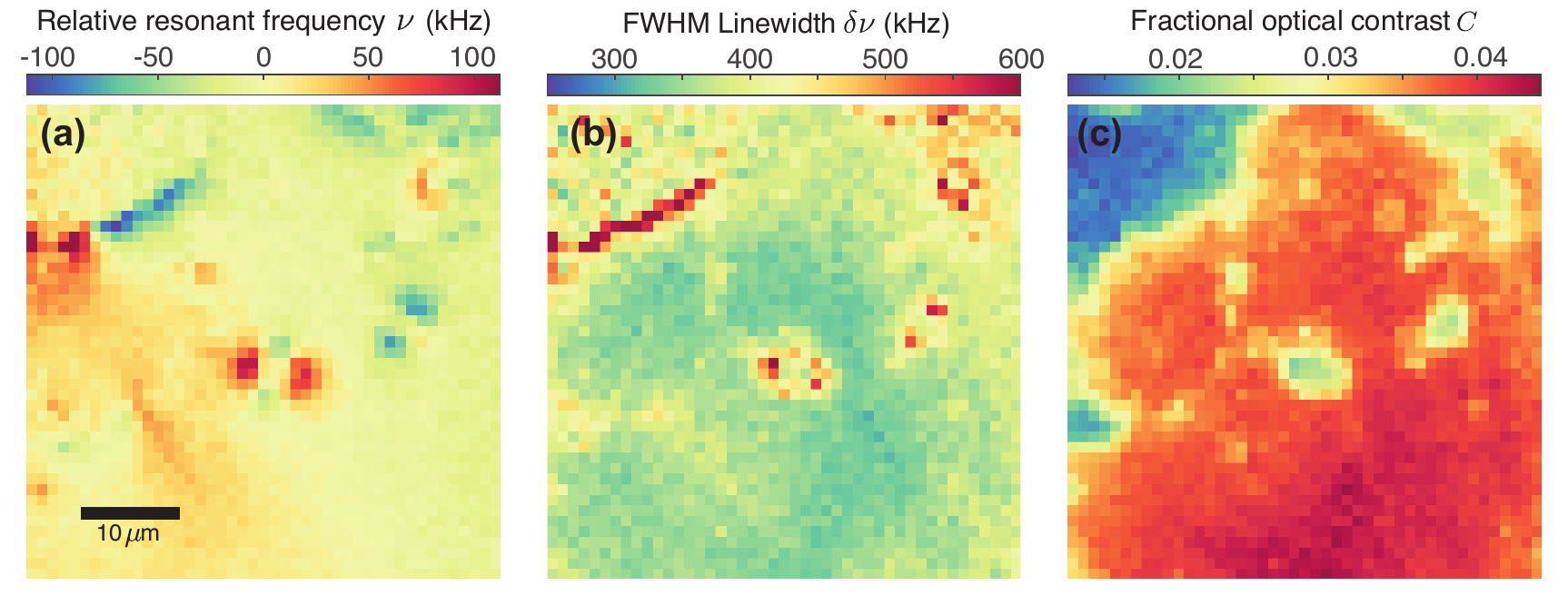}
   \centering
\caption{\textcolor{black}{Multiple mechanisms lead to changes in NV ODMR on a micron scale when imaging static magnetic dipole and strain fields. (a) Resonant frequency $\nu$ shifts due to magnetic field and crystal strain. (b) The FWHM linewidth $\delta \nu$ varies due to \textcolor{black}{gradients of magnetic and strain fields that inhomogenously broaden the NV ODMR in each pixel. (c) Optical contrast $C$ also varies due to due to inhomogenous broadening of the NV ODMR.}}}
\label{fig:sqcandlw}
\end{figure*}

\section*{Static Magnetic Imaging Modality}

\textcolor{black}{To measure static fields,} the full NV resonance curve can be measured in wide-field with arbitrarily long acquisition times. Taking a series of PL images over a range of RFs allows fitting of the entire resonant response in the measured range. Mapping the fitted resonant frequency at each pixel results in a partial reconstruction of the magnetic field as seen in Fig.~\ref{fig:sqcandlw}(a). Taking the difference between the $|m_s=0\rangle\leftrightarrow|m_s = \pm1\rangle$ resonant frequency maps eliminates shifts of the resonance due to fields other than the magnetic field, and enables imaging of the absolute magnetic field projection along the NV symmetry axis~\cite{Fescenko2019DiamondNanocrystals}, as seen in Fig.~\ref{fig:DDQcomparison}(a). 

For quickly-varying magnetic fields, \textcolor{black}{imaging via the static magnetic imaging procedure} may not be possible. We thus require a dynamic imaging modality that reproduces the absolute magnetic field across the imaging field of view \textcolor{black}{without prior per-pixel calibration} and is compatible with high-frame-rate imaging.

\section*{Dynamic Magnetic Imaging Modalities}

\subsection*{Single Quantum Difference Imaging}

The simplest dynamic imaging modality uses RF excitation applied at one inflection point of \textcolor{black}{one of} the \textcolor{black}{NV} resonance curves [Fig.~\ref{fig:NVexplanation}(e)]. PL images taken with RF $\pi$-pulses applied are subtracted from PL images taken without RF to detect changes in emitted NV PL ~\cite{Wojciechowski2018Camera-limitsSensor}. We define a SQ difference image (DI) as 
\begin{equation}
    \text{SQ}(f_1) = \frac{I_{\text{off}} - I_{\text{on}}(f_1)}{I_{\text{off}}},
    \label{equation:sqdi}
\end{equation}
in which $I_{\text{on}}(f_1)$ is the intensity image taken with applied RF $\pi$-pulses at $f_1$ and $I_{\text{off}}$ is the image taken with no applied RF. \textcolor{black}{We assume the NV ensemble is operating in the linear-response regime of the resonance curve, i.e. $\gamma_{\text{NV}}\hat{z}_{\text{NV}}\cdot\vec{B} < \delta \nu$. The per-pixel signal is then given by}
\begin{equation}
    \text{SQ}^{\text{pp}}(f_1) = \frac{9}{8} C - \frac{3 \sqrt{3}}{4} \frac{C}{\delta\nu} (\nu(\vec{E}, \vec{B}, T, ...) - f_1).
    \label{equation: single quantum difference signal}
\end{equation}
As discussed above, the resonant frequency $\nu$ depends on the magnetic field $\vec{B}$ and also varies with local electric field $\vec{E}$, temperature $T$, and crystal strain. 

\begin{figure*}
    \begin{center}
     \includegraphics[width=1.8\columnwidth]{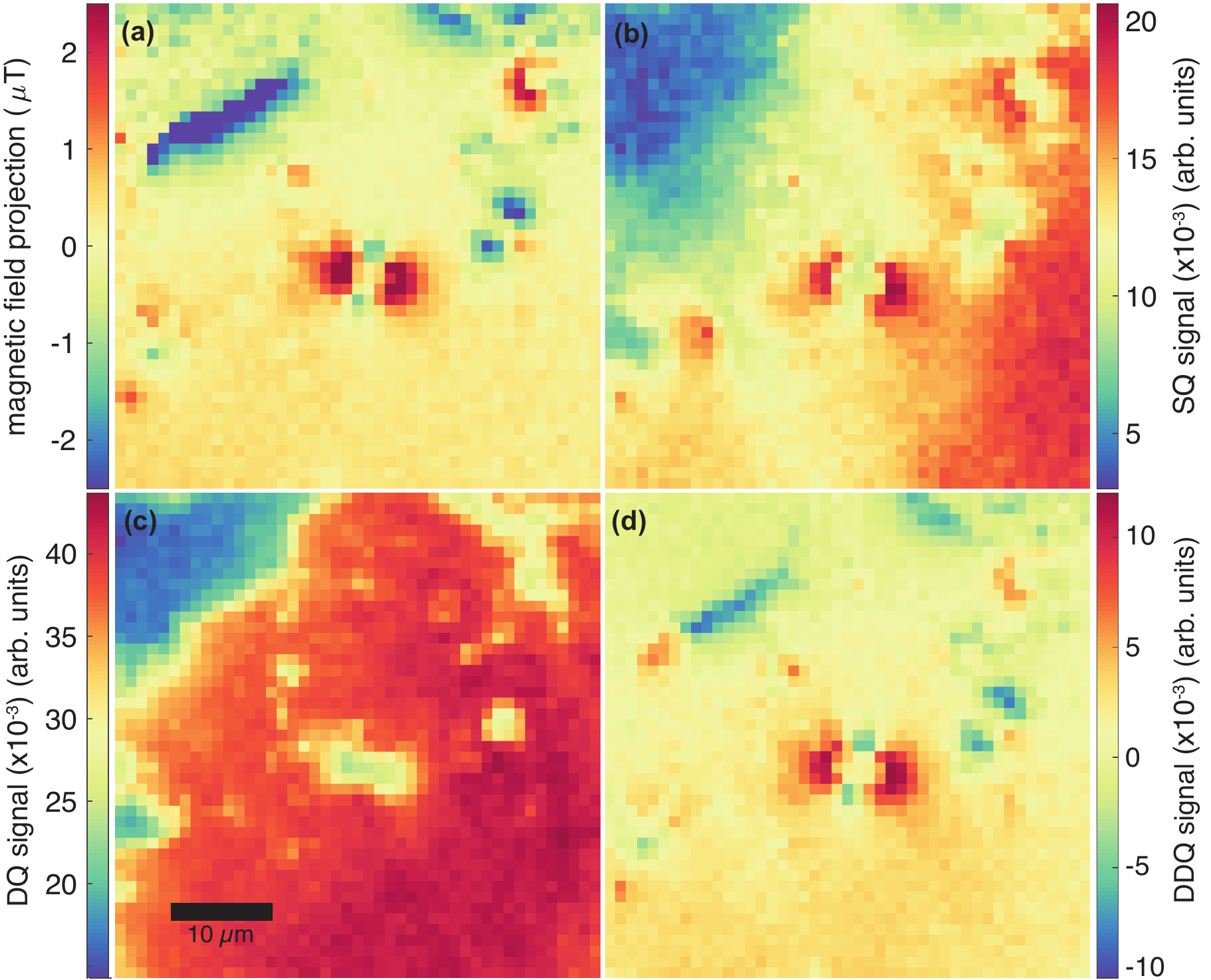}
     \end{center}
\caption{\textcolor{black}{Competing magnetic imaging modalities. (a) \textcolor{black}{`True'} static magnetic field projection map generated with the frequency scanning technique outlined in the above Static Magnetic Imaging Modality section (acquisition time \SI{12}{\second}). (b) Single quantum difference imaging (\SI{2.4}{\second}). The signal measured with the SQ DI modality is a convolution of the magnetic and strain fields, which are impossible to separate with a single measurement. (c) Double quantum difference imaging (\SI{2.4}{\second}). While the DQ DI modality has reduced the impact from the fields that homogeneously shift the NV centers, the DQ signal is more sensitive to the local contrast and linewidth variations of the NV sensing curves. (d) Double-double quantum difference imaging (\SI{2.4}{\second}). An inspection of the competing dynamic imaging schemes (b-d) reveals that both the SQ (b) and DQ (c) schemes are significantly compromised by spurious contrast caused by strain gradients and curve-shape variation, respectively, while the DDQ scheme (d) faithfully approximates the \textcolor{black}{`true'} magnetic field projection (a) with a decreased acquisition time.}}
\label{fig:DDQcomparison}
\end{figure*}

Additionally, variations in curve-shape, and thus $C$ [Fig.~\ref{fig:sqcandlw}(b)] and $\delta\nu$ [Fig.~\ref{fig:sqcandlw}(c)], cause variations of the resonance curve inflection point, contributing to the SQ signal. Microscopic and mesoscopic strain inhomogeneities result in inhomogeneous broadening of the resonance curve~\cite{Kehayias2019ImagingCenters}. Also, dephasing due to dipolar interactions between \textcolor{black}{sensing} NV centers and other paramagnetic impurities (e.g. P1 and NV) fundamentally limit the NV ensemble coherence, and thus, ODMR linewidth~\cite{Kleinsasser2016,Bauch2018UltralongControl}. 

The SQ DI enables imaging of some nano-scale magnetic structure, but the sensitivity of this technique is limited. In Fig.~\ref{fig:DDQcomparison}(b) we show a SQ DI image with its corresponding static magnetic field map in \ref{fig:DDQcomparison}(a). The SQ DI enables partial mapping of the magnetic field projection, \textcolor{black}{but is limited by contributions to the resonant frequency shift by strain gradients, as seen in the SQ-signal-color-gradient from upper-left to lower-right in Fig.~\ref{fig:DDQcomparison}(b).}

\subsection*{Double Quantum Difference Imaging}

Temperature, electric field, and strain shift the zero-field-splitting of the NV ground state, causing common-mode shifts of the $|m_s=0\rangle~\leftrightarrow~|m_s~=~\pm~1\rangle$ transitions~\cite{Levine2019PrinciplesMicroscope}. Conversely, the magnetic-field-induced Zeeman effect splits the two transitions. Thus, by probing the difference of the two transition resonant frequencies, the common-mode shifts can be subtracted out and the magnetic field projection can be measured directly. We use a DQ driving scheme~\cite{Fang2013High-SensitivityCenters,Myers2017Double-QuantumCenters}, applying two-tone RF $\pi$-pulses at the opposite inflection points of the two resonance curves simultaneously [Fig.~\ref{fig:NVexplanation}(e)]. We construct a DQ DI by subtracting the PL image taken with DQ RF driving from an image taken with no RF applied
\begin{equation}
    \text{DQ}(f_{1},f_{4}) = 
    \frac{I_{\text{off}} - I_{\text{on}}(f_{1},f_{4})}{I_{\text{off}}},
\end{equation}
in which $I_{\text{on}}(f_1,f_4)$ is the image taken with applied RF $\pi$-pulses \textcolor{black}{at $f_1$ and $f_4$ simultaneously}, and $I_{\text{off}}$ is the image taken with no applied RF. \textcolor{black}{We again assume linear-response of the resonance curves and additionally assume that the two resonance curves have the same shape \textcolor{black}{(see Supplemental Material)}}. The per-pixel DQ signal is
\begin{multline}
\text{DQ}^{\text{pp}}(f_1, f_4) \approx \frac{9}{4} C - \frac{3 \sqrt{3}}{4} \frac{C}{\delta\nu}(f_{1} - f_{4}) 
  \\ - \frac{3 \sqrt{3}}{4} \frac{C}{\delta\nu} (2 \gamma_{\text{NV}} \hat{z}_{\text{NV}} \cdot \vec{B}).
  \label{equation: unsimplified DQ linear signal}
\end{multline}
By defining $\langle\vec{B}\rangle$ as the average magnetic field over the imaging field of view, Eq.~\ref{equation: unsimplified DQ linear signal} simplifies to
\begin{multline}
\text{DQ}^{\text{pp}}(f_1, f_4) = \frac{9}{4} C + \frac{3 \sqrt{3}}{4} \frac{C}{\delta\nu}2 \delta_0 
  \\ - \frac{3 \sqrt{3}}{4} \frac{C}{\delta\nu} \left( 2 \gamma_{\text{NV}} \hat{z}_{\text{NV}} \cdot \left(\vec{B} - \langle\vec{B}\rangle \right) \right),
  \label{equation: double quantum difference signal}
\end{multline}
where $2\delta_0 = (f_4-f_1)-2\gamma_{\text{NV}}\hat{z}_{\text{NV}}\cdot\langle\vec{B}\rangle$. By applying $f_1$ and $f_4$ at the outer inflection points of the NV resonance curves \textcolor{black}{simultaneously} [Fig.~\ref{fig:NVexplanation}(e)], intensity changes induced by non-magnetic, common-mode shifts are cancelled out, while splittings caused by magnetic signal result in a sum of changes in PL intensity. Hence, for constant $C$ and $\delta\nu$, the DQ DI technique enables absolute magnetic imaging.~\cite{Fang2013High-SensitivityCenters} 


\textcolor{black}{Although the contribution of strain-induced-resonance-shifts to the imaging have been eliminated, overcoming the sensitivity limits of the SQ DI, we demonstrate that} DQ DI has \textit{increased} the effect of variations in curve-shape on the magnetic imaging. \textcolor{black}{More specifically,} variations of $C$ and $\delta\nu$ still cause perturbations of the first two terms in Eq.~\ref{equation: double quantum difference signal}. This effect can be seen by comparing the map of $C$ in Fig.~\ref{fig:sqcandlw}(c) to the DQ DI in Fig.~\ref{fig:DDQcomparison}(c). \textcolor{black}{Thus, for practical applications of the DQ method to wide-field imaging, we find that curve-shape variations dominate and the DQ DI (Fig.~\ref{fig:DDQcomparison}(c)) is ineffective at reproducing a map of the magnetic field projection (Fig.~\ref{fig:DDQcomparison}(a)).}
\\
\\
\newline
\subsection*{Double-Double Quantum Difference Imaging}
\textcolor{black}{To suppress the imaging dependence on curve-shape, we apply bias RFs on either side of the resonance curves \cite{Gould2014}.} We construct a DDQ DI
\begin{equation}
         \textcolor{black}{\text{DDQ} = 2\frac{
         I_{\text{on}}(f_1,f_4) - I_{\text{on}}(f_2, f_3)}
         {I_{\text{on}}(f_1,f_4) + I_{\text{on}}(f_2, f_3)}
         }
         \label{equation: ddq from images signal}
\end{equation}
where $I_{\text{on}}(f_1, f_4)$ ($I_{\text{on}}(f_2, f_3)$) is the image taken with RF applied at the outer (inner) inflection points of the two resonance curves \textcolor{black}{simultaneously}, as shown in Fig.~\ref{fig:NVexplanation}(e). \textcolor{black}{By applying DQ bias RFs on either side of the resonance curves, the effects of variations in the shape of the the ODMR curve and external non-magnetic fields are mitigated. Here, the DDQ DI signal is normalized by dividing by the mean of the individual DQ frames $I_{\text{on}}(f_i,f_j)$.}
 
\begin{figure*}
    \begin{center}
     \includegraphics[width=2.1\columnwidth]{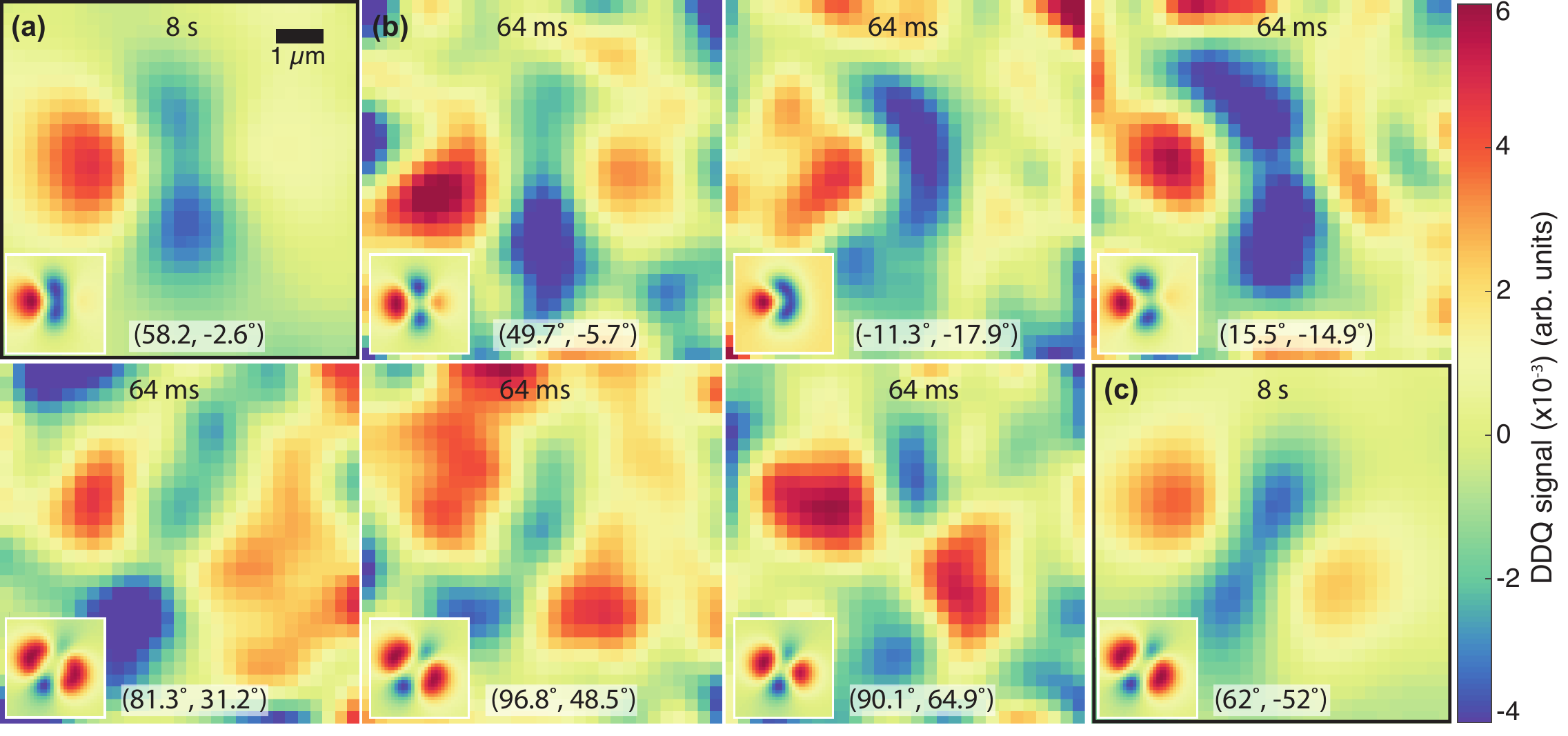}
     \end{center}
\caption{\textcolor{black}{DDQ imaging of the reorientation of a DNA-tethered magnetic nanoparticle under applied flow. In each panel, the observed DDQ image is compared with a \textcolor{black}{fitted DDQ image (inset)} to estimate the magnetic nanoparticle dipole orientation ($\theta$,~$\phi$), where the NV ensemble symmetry axis is (54.735\degree,~0\degree). For all DDQ DI in this figure, a Gaussian smoothing filter with $\sigma = $ \SI{533}{\nano\meter} is applied. (a) A time-averaged DDQ DI (\SI{8}{\second}) showing initial magnetic nanoparticle orientation before flow. (b) Representative DDQ frames (\SI{64}{\milli\second} of exposure) showing nanoparticle reorientation in response to an applied flow. (c) A time-averaged DDQ DI (\SI{8}{\second}) showing the final magnetic nanoparticle orientation with applied flow.}} 
\label{fig:dynamics}
\end{figure*}

Because of the choice of RF, the per-pixel DDQ signal simplifies in a similar manner as the DQ signal in Eq.~\ref{equation: double quantum difference signal} giving
\begin{equation}
  \text{DDQ}^{\text{pp}} \approx \frac{3 \sqrt{3}}{2} \frac{C}{\delta\nu}\left(2 \gamma_{\text{NV}} \hat{z}_{\text{NV}} \cdot \left(\vec{B} - \langle\vec{B}\rangle\right) \right).
  \label{equation: double double quantum difference signal}
\end{equation} 
DDQ eliminates the first two terms of the DQ DI signal Eq.~\ref{equation: double quantum difference signal} to obtain a single term which is linearly proportional to $(\vec{B} - \langle \vec{B} \rangle)$. There is still multiplicative dependence on $C$ and $\delta\nu$, but because the shift of the resonant frequency far \textcolor{black}{($>$\SI{1}{\micro\meter})} from magnetic field sources falls off faster than the impact of spatial variations of curve-shape \textcolor{black}{due to inhomogeneous broadening}, there is no \textcolor{black}{DDQ signal} generated in regions with no magnetic field. The DDQ \textcolor{black}{DI} completely eliminates the large-scale, non-magnetic gradients in the SQ DI [Fig. $\ref{fig:DDQcomparison}$(b)] and suppresses the $C$ and $\delta\nu$ dependence [Fig.~\ref{fig:sqcandlw}(b-c)] of the DQ DI [Fig.~\ref{fig:DDQcomparison}(c)]. As shown in Fig.~\ref{fig:DDQcomparison}(d), DDQ DI provides similar magnetic sensitivity as the static magnetic projection map in Fig.~\ref{fig:DDQcomparison}(a), with a \textcolor{black}{greater than four}-fold \textcolor{black}{acquisition-time-reduction}. The static imaging modality requires enough images to fit both resonance curves; the DDQ modality instead extracts the magnetic field dependence of the resonances with only two images: $I_{\text{on}}(f_1,f_4)$ and $I_{\text{on}}(f_2,f_3)$. While the integration time of the DDQ image shown in Fig.~\ref{fig:DDQcomparison}(d) was chosen to match the signal-to-noise ratio of the magnetic field map in Fig.~\ref{fig:DDQcomparison}(a), DDQ enables even faster magnetic imaging \textcolor{black}{as demonstrated in the next section}.

\textcolor{black}{We emphasize the general conditions for applicability for the DDQ method: (i) the resonance curve shapes of the two NV spin transitions used must be matched by driving each transition with equal Rabi frequency, (ii) non-magnetic inhomogeneities across the imaging field of view must be smaller than the resonance FWHM linewidth in order to be suppressed, and (iii) sensor operation is in the linear regime, i.e. magnetic signals to be imaged are smaller than the resonance FWHM linewidth. We discuss errors associated with condition (i) in the Supplemental Material, and note criteria (ii) and (iii) are prerequisites for any intensity-based wide-field magnetic imaging involving NV ensembles.}

\textcolor{black}{
\section*{WIDE-FIELD DYNAMIC MAGNETIC MICROSCOPY}
To demonstrate that DDQ difference imaging can facilitate high-frame-rate imaging of dynamic fields, we image the changing magnetic field produced by a ferromagnetic nanoparticle tethered to the diamond sensor surface by a single DNA molecule. The approximate diamond-particle distance is \SI{400}{\nano\meter}. Fig.~\ref{fig:dynamics}a shows an \SI{8}{\second} time-averaged image of the field produced by the magnetic nanoparticle. A preferred orientation is observed due to the partial alignment of the ferromagnetic nanoparticle to the \SI{0.35}{\milli \tesla} external magnetic-field, oriented along $(\theta,\phi)=(54.735\degree,0\degree)$. Next, phosphate-buffered-saline is pulled from a reservoir through the sample channel by a syringe pump at 4 ml/min. The fluid flow exerts a hydrodynamic force and torque on the tethered-particle, causing it to reorient, changing the magnetic field at the diamond sensor surface. Fig.~\ref{fig:dynamics}b displays characteristic frames, in chronological order, showing time-resolved imaging of the nanoparticle moment reorientation at a \SI{15.6}{\hertz} frame-rate \textcolor{black}{(\SI{64}{\milli \second} per frame)}, with insets showing \textcolor{black}{fitted} DDQ images displaying the changing magnetic-moment direction in each frame. The fluid-flow-steady-state nanoparticle orientation is imaged with an \SI{8}{\second} time-averaged DDQ image in Fig.~\ref{fig:dynamics}c. The full dynamic magnetic imaging video \textcolor{black}{and simulation information} can be found in the Supplemental Material. This experiment represents the novel application of micron-scale dynamic magnetometry to a single-molecule biological system.
}

\section*{Conclusion and Outlook}
\textcolor{black}{Although} the NV community has made significant progress toward eliminating inhomogeneities in NV ensemble-based-sensors \textcolor{black}{through advanced NV fabrication}~\cite{Acosta2009DiamondsApplications}, quantum control methods \textcolor{black}{can significantly increase the sensitivity of these systems for magnetometry applications}~\cite{Bauch2018UltralongControl}. \textcolor{black}{However, existing wide-field schemes fail to reliably image magnetic fields due to micron-scale variation in the resonance-curve shape.} Here, we introduce a novel quantum control technique, double-double quantum difference imaging, that is suitable for mitigating inhomogeneities in wide-field dc magnetometry to enable imaging of time-varying fields. Using four-tone RF pulses and only a two-image sequence, we show both theoretically and experimentally that DDQ difference imaging not only mitigates \textcolor{black}{non-magnetic} perturbations of the NV resonant frequency but also variations of resonance curve-shape. \textcolor{black}{Static-field imaging reveals} that these resonance shape variations can be the dominant source of imaging noise in a state-of-the-art NV magnetic imaging surface. \textcolor{black}{Finally, we use the DDQ technique to perform wide-field magnetic microscopy of a dynamic, biological system, enabling high frame-rate orientation imaging of a magnetic nanoparticle tethered to the diamond sensor by a single DNA molecule. DDQ difference imaging eliminates the need for per-pixel calibration and enables high-frame-rate magnetic microscopy via NV photoluminescence intensity imaging.}
\newline
\newline
\textcolor{black}{\textit{During the revision process, we became aware of work demonstrating a similar technique with similar conditions of applicability to mitigate wide-field inhomogeneities in NV magnetic microscopy~\cite{Hart2020}. In relation to this work, our technique utilizes a simpler two-image (versus four-image) scheme, requires no phase control of the RF excitation, uses substantially lower RF power, and was used to perform dynamic magnetic imaging.}}

\section*{Acknowledgements}
This material is based on work supported by the National Science Foundation under Grant No. 1607869. Helium ion implantation measurements were carried out at the Environmental and Molecular Sciences Laboratory, a national scientific user facility sponsored by DOE?s Office of Biological and Environmental Research and located at Pacific Northwest National Laboratory (PNNL). PNNL is a multiprogram laboratory operated for DOE by Battelle under Contract DE-AC05-76RL01830. K.M.I. acknowledges support from the Spintronics Research Network of Japan.

\end{document}